\documentclass[aps,prl,twocolumn,superscriptaddress]{revtex4}

\usepackage{graphicx}
\usepackage[english]{babel}
\newcommand{\ang}{\ensuremath{\text{\AA}}}
\newcommand{\diff}{\ensuremath{\text{d}}}

\newcommand{\Tr}{\,{\rm{Tr}\,}}

\newcommand{\sign}{\,{\rm{sign}\,}}

\renewcommand{\Im}{{\rm Im}}

\newcommand{\beqa}{\begin{eqnarray}}
\newcommand{\eeqa}{\end{eqnarray}}

\begin{document}

\title{Phonon mediated tunneling into graphene}

\author{T. O. Wehling}
\affiliation{I. Institut f{\"u}r Theoretische Physik, Universit{\"a}t Hamburg, Jungiusstra{\ss}e 9, D-20355 Hamburg, Germany}
\author{I. Grigorenko}
\affiliation{Theoretical Division, Los Alamos National Laboratory, Los Alamos, New Mexico 87545,USA}
\author{A. I. Lichtenstein}
\affiliation{I. Institut f{\"u}r Theoretische Physik, Universit{\"a}t Hamburg, Jungiusstra{\ss}e 9, D-20355 Hamburg, Germany}
\author{A. V.  Balatsky}
\email[]{avb@lanl.gov, http://theory.lanl.gov}
\affiliation{Theoretical Division, Los Alamos National Laboratory, Los Alamos, New Mexico 87545,USA}
\affiliation{Center for Integrated
Nanotechnologies, Los Alamos National Laboratory, Los Alamos, New
Mexico 87545,USA}

%\thanks{}

\date{\today}

\begin{abstract}
Recent scanning tunneling spectroscopy experiments\cite{brar_STS1, Crommie_STS} on graphene reported an unexpected gap of about $\pm 60$\,meV around the Fermi level. Here, we give a theoretical investigation explaining the experimentally observed spectra and confirming the phonon mediated tunneling as the reason for the gap: We study the real space properties of the wave functions involved in the tunneling process by means of ab-initio theory and present a model for the electron-phonon interaction, which couples the graphene's Dirac electrons with quasi free electron states at the Brillouin zone center. The self-energy associated with this electron-phonon interaction is calculated and its effects on tunneling into graphene are discussed. In particular, good agreement of the tunneling density of states within our model and the experimental\cite{brar_STS1, Crommie_STS} $\diff I/\diff U$ spectra is found.
\end{abstract}

% \pacs{68.37.Ef, 71.55.-i, 81.05.Uw}
%\keywords{}

\maketitle

%\section{Introduction}
Graphene, the two dimensional allotrope of carbon
\cite{Novoselov_science2004}, is famous for its electrons having
vanishing effective mass \cite{Geim2005,Zhang2005}. At two
non-equivalent points at the corner of the Brillouin zone, K and K',
the linearly dispersing valence and conduction band touch. So,
electrons in graphene behave like massless Dirac fermions with the
speed of light being replaced by the Fermi velocity $v_{\rm
f}\approx c/300$. This yields an interesting analogy between
condensed matter and high energy physics \cite{KatsQED}, which has
been extensively studied during the last years\cite{AHC2007}. We
argue, however, that there is a decisive difference between the
Dirac fermions in graphene and Dirac particles studied in the
context of high energy physics: The electrons in graphene are
embedded to a real material which interacts with these electrons.
This "real material background" has proven to introduce spatial
charge inhomogeneities due to the imperfections present in any solid
\cite{MartinSEM}: The imperfections include charged impurities or
structural corrugations. Neither free standing graphene nor graphene
on a substrate is perfectly flat \cite{meyer2007,fasolino}, which
leads to effective gauge fields acting on the Dirac electrons
\cite{morpurgo}.

Recently reported scanning tunneling
spectroscopy (STS) experiments on graphene
stress this real material backround even more
 \cite{brar_STS1,Crommie_STS}. It is not the
 Dirac point with its linearly vanishing density of density
  of states which causes the most prominent feature
in the measured $\diff I/\diff U$ spectra,
but a gap of $\pm 60$\,meV pinned to the
Fermi level \cite{Crommie_STS}. After initial
speculations on the nature of this gap like
feature including substrate and electric field
effects \cite{brar_STS1}, the most recent experiments
 indicate that this gap is caused by the opening of an
  inelastic tunneling channel due to graphene's out-of-plane
  at K and K' \cite{Crommie_STS}.

Here, we give a detailled theoretical study of the
latter scenario. Firstly, we investigate the effect
of the K/K' out-of-plane phonons on the electronic
wave functions and their decay in the vacuum by means
of first principles theory. We show why the electron-phonon
 coupling has  a large impact on STS as seen in Refs.
  \cite{brar_STS1, Crommie_STS}. Motivated by this insight,
  we present a simple model of graphene's electrons being
   coupled to the out-of-plane phonons. Within this model the
   electron self-energy, the total (DOS) and the tunneling density of
   states (TDOS) are calculated. It turns out that the total DOS and the TDOS, which is significant for STS, differ strongly: At low energies, the total DOS is dominated by the $V$-shape from the Dirac electron's DOS. However, the TDOS recovers the gapped experimental $\diff I/\diff U$ spectra. Here, the inelastic channel is strongly enhanced in the TDOS due to a very general mechanism of band mixing. This mechanism is not limited to dynamic processes but is also expected to occur, e.g., near short range corrugations of the graphene lattice.

To address the effects of the K/K' phonons on the electronic wave functions of graphene, we performed density functional theory calculations within the framework of the local density approximation using the Vienna Ab Initio Simulation Package (VASP) \cite{Kresse:PP_VASP} with the projector augmented wave (PAW) \cite{Kresse:PAW_VASP,Bloechl:PAW1994} basis sets. The corresponding plane wave expansions were cut-off at $928$\,eV and the Brillouin zone integrations were carried out with the tetrahedron-method on k-meshes denser than $20\times 20$ when folded back to the simple graphene unit cell. The vertical extension of this cell was chosen to be $24\ang$.

\begin{figure}
\begin{minipage}{.92\linewidth}
\includegraphics[width=\linewidth]{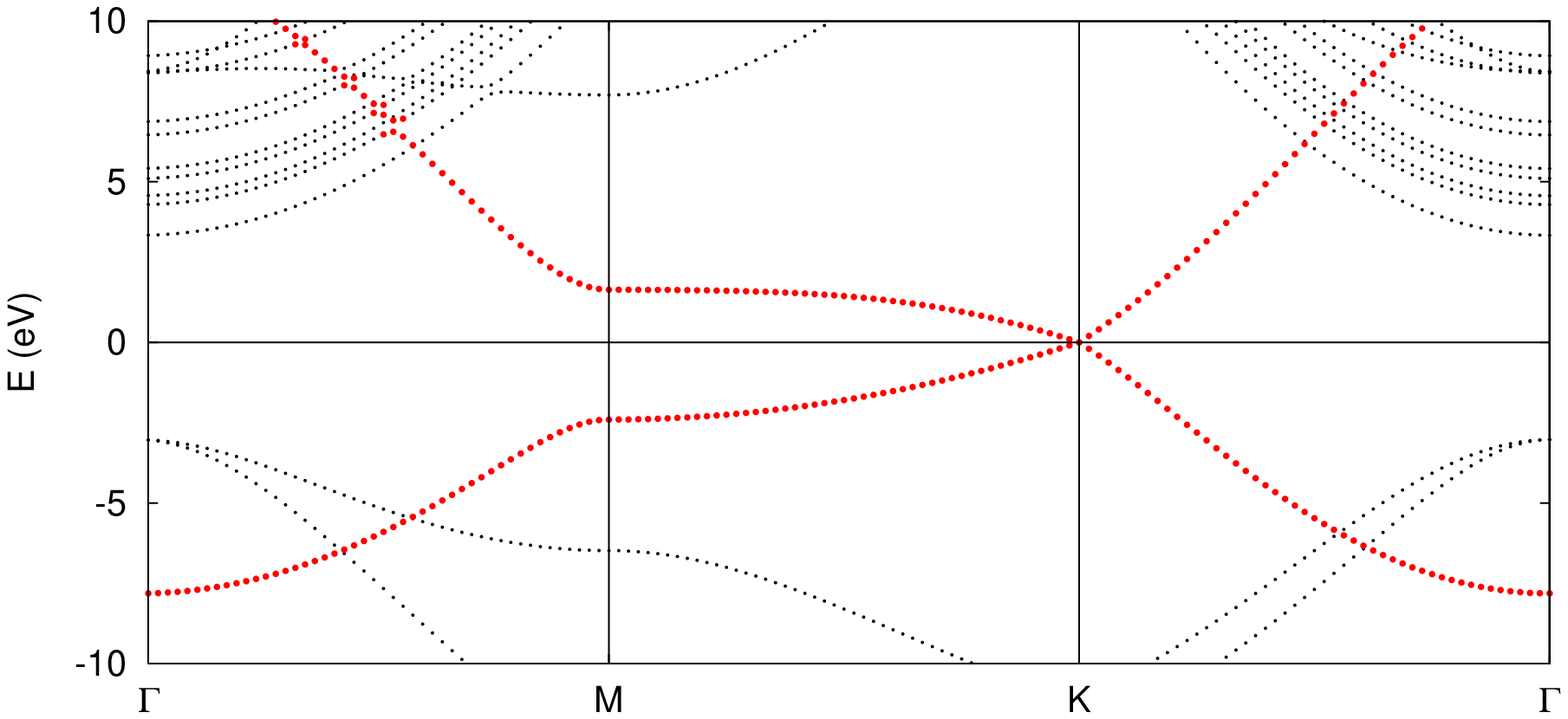}\vspace{.2cm}
\end{minipage}
\begin{minipage}{.9\linewidth}
\includegraphics[width=\linewidth]{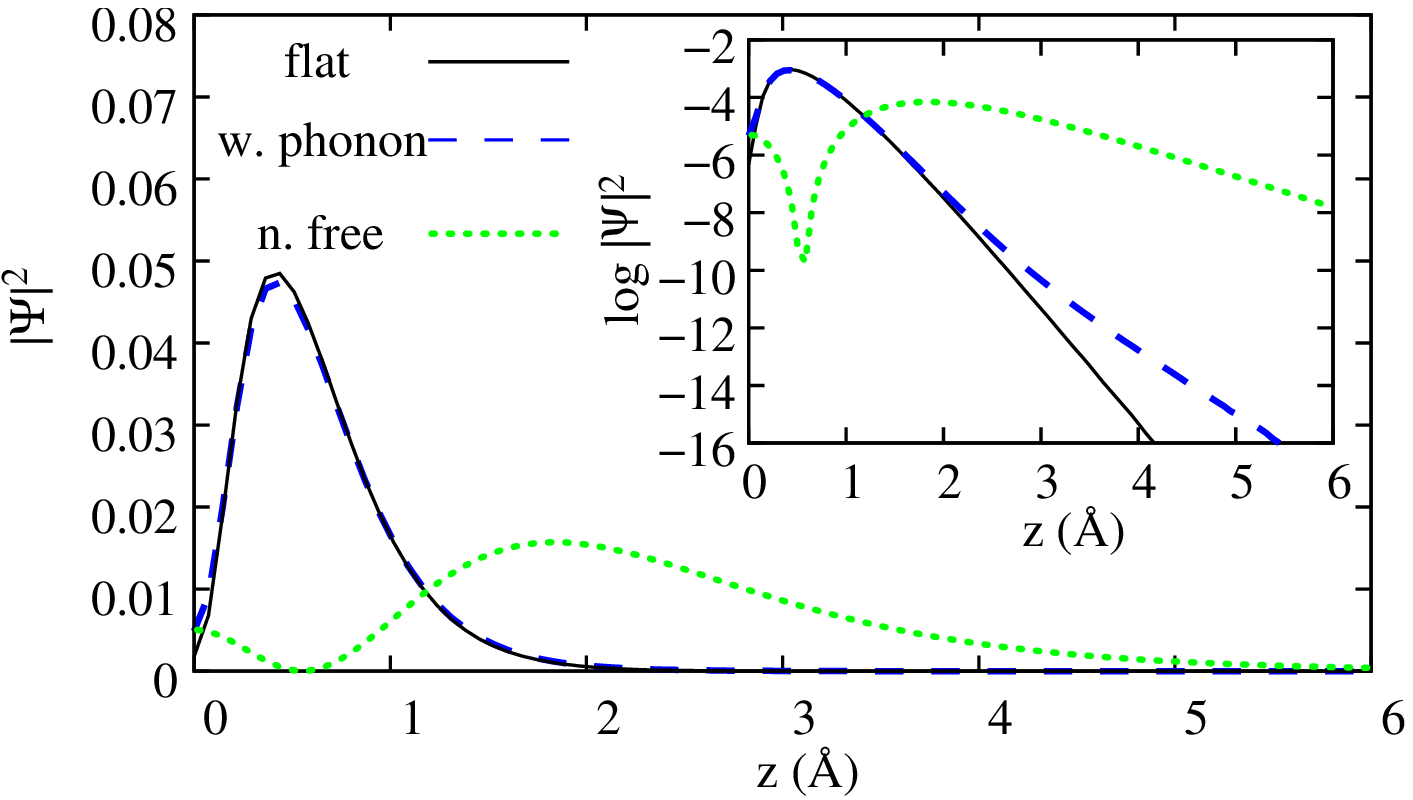}
\end{minipage}
\caption{\label{fig:bands_decay}(Color online)  Upper panel: Band structure of graphene. The $\pi$-bands inhibiting the Dirac electrons are marked with fat red dots. Instead of the graphite interlayer band, in graphene a quasi-continuum of nearly free electron states begins $3.3$\,eV above the Fermi level at the $\Gamma$-point. Lower panel: Decay of the electronic wave functions $\Psi(z)$ as a function of hight $z$ above the graphene sheet. The laterally averaged density $|\Psi(z)|^2$ is shown for wave functions near the Dirac point of flat graphene (solid line) and in presence of the frozen phonon (dashed line). The density belonging to the lowest quasi free electron band at $\Gamma$ printed as dotted line.}
\end{figure}

The band structure of flat single layer graphene obtained in this way is shown in the
upper part of Fig. \ref{fig:bands_decay}. The $\pi$-bands (fat red dots) exhibit the
well known shape intersecting the Fermi level at the K point with linear dispersion
in the vicinity. At the zone center $\Gamma$, a quasi continuum of free electron like
 bands starts $3.3$\,eV above the Fermi level. These
 bands  correspond to the interlayer band of graphite, which plays a
  crucial role in graphite intercalation compounds \cite{Cambridge2005}. In
  contrast to graphite, the quasi free electron bands, here, are not limited
   by other graphene layers on top and extend far into the vacuum above the sheet.

To quantify the decay of the wave functions of different bands in the vacuum, we calculated the laterally averaged charge density $|\Psi(z)|^2$ of these wave functions as a function of distance $z$ from the sheet. As shown in Fig. \ref{fig:bands_decay}, lower panel, the wave functions at the Dirac point of flat graphene have their maximum at $z=0.4\;\ang$ above the sheet. For $z\gtrsim 1.5\ang$, their probability density starts decaying into the vacuum dominated by an exponential with decay constant $\lambda_K^{-1}=3.9\ang^{-1}$, which has been obtained from a linear fit of $\log|\Psi_K(z)|^2$ in this region. This decay constant follows from the single particle Schr{\"o}dinger equation in presence of a constant potential given by the graphene work function \cite{TersoffHamann}: $\lambda_\pi^{-1}=2\sqrt{k_{||}+2m_{\rm f}\phi/\hbar^2}$, where the in-plane crystal momentum is $|k_{||}|=1.7\ang^{-1}$ for the electrons at K/K' and we obtained $\phi=4.5$\,eV for the work function from LDA.

%\section{Method}
States at the $\Gamma$ point of the Brillouin zone, $k_{||}=0$, decay much slower
 into the vacuum: The lowest nearly free electron state is extremely long ranged
 (see Fig 1., lower panel) and spreads out into the vacuum with decay constant
  $\lambda^{-1}=1.0\ang^{-1}$. Near the Fermi level $E_{\rm f}$, slowly decaying
  electronic states are generated by out-of-plane K- and K'-phonons, which mix
  the nearly free electron bands at $\Gamma$ with the Dirac-like $\pi$ bands at K.
To illustrate this point, consider a frozen phonon with
amplitude %$\Delta z=\hbar/(2M_C\omega_a)$
$\Delta z=0.1\ang$ and a wave vector connecting K/K'.
In presence of this phonon the Dirac bands formerly at
K and K' are folded back to the $\Gamma$ point of the
$(\sqrt{3}\times\sqrt{3})$\,R$30^\circ$ super cell
 inhibiting the frozen phonon. One out of the Dirac
 bands touching the Fermi level at $\Gamma$ hybridizes
 with the long range quasi free electron states upon
 formation of the frozen phonon. This results in the
 long range decay of the corresponding charge density,
  $|\Psi_\Gamma(z)|^2\sim e^{-\lambda_\Gamma^{-1}z}$
  with $\lambda_\Gamma^{-1}=2.2\ang^{-1}$, being approximately
   twice slower than for the unperturbed Dirac bands.

Experimentally, the decay of the wave functions
 involved in the tunneling processes has been obtained
 from the dependence of the tunneling current on the
  tip-sample distance\cite{Crommie_STS}. Different
  decay constants $\lambda_{\rm in}^{-1}\approx 4\ang^{-1}$
  and $\lambda_{\rm out}^{-1}\approx 2\ang^{-1}$ were measured
  inside and outside the gap \cite{Crommie_STS}, respectively,
  which nicely coincide with the theoretical values
  $\lambda_K^{-1}$ and $\lambda_\Gamma^{-1}$.

According to Tersoff-Hamann theory, tunneling currents are
determined by the tunneling density of states, which is the LDOS at
the location of the tip - usually on the order $5\ang$ above the
sample. Thus, for each band, the DOS has to be weighted with squared
amplitude of the corresponding wave function about $5\ang$ above the
sample - which can be approximately written as $\diff I/ \diff U
\sim |\Psi_\Gamma|^2N_\Gamma(E)+|\Psi_K|^2N_K(E)$. With the decay constants from above for the Dirac bands of flat graphene and in presence of the phonon, the ratio of the exponential prefactors entering the tunneling matrix elements is $|\Psi_\Gamma/\Psi_K|^2\propto e^{1.7\ang^{-1} z}$. At $z=5\ang$ this ratio is on the order of $10^4$. Thus, the effect of mixing of the free electron bands and the $\pi$-bands by electron-phonon coupling on STS spectra will be greatly enhanced by this factor of $10^4$.

In the following, we present a simple model to understand how electron-phonon coupling due to the K/K' out of plane phonons affects STS: As the previous discussion showed, most important for these experiment will be the mixing of the quasi free electron bands at $\Gamma$ with the Dirac fermion bands at K/K'. The former bands have their minimum $E_\sigma=3.3$\,eV above the Fermi level and may be approximated by a flat band $H_\sigma=\sum_{q} d^\dagger_q E_\sigma d_q$ in the vicinity of $\Gamma$. Here, $d_q$ is the annihilation operator of an electron with crystal momentum $q$ in the nearly free electron band.
The two $\pi$ bands give rise to graphene's linear density of states close to the Dirac point and their Hamiltonian $H_\pi$ may be written as $H_\pi=\sum_{\nu=\pm,q}\nu\epsilon(q) c^\dagger_{\nu,q}c_{\nu,q}$, where the index $\nu=+(-)$ denotes the conduction (valence) band and $c_{\nu,q}$ is the annihilator of an electron in this $\pi$-band with momentum q. Close to the Dirac points K and K', i.e. $q=\pm K+k$ with $|k|\ll|K|$, the dispersion is linear $\epsilon(k)\approx\hbar v_{\rm f}|k|$.

Out-of-plane phonons scatter electrons between the $\pi$ and quasi free electron bands with the electron-phonon-interaction reading as
\begin{equation}
\label{eqn:V_e-ph}
V=\lambda \sum_{\nu,q,k} (d^\dagger_{k+q}c_{\nu,q}+c^\dagger_{\nu,k+q} d_{q})(a_k+a^\dagger_{-k}),
\end{equation}
where $a_k$ annihilates an out-of-plane phonon carrying crystal momentum $k$.
As stated above, the phonon modes at K and K' will be the most important contributors to inelastic tunneling signals. Around these points, their Hamiltonian can be approximated by $H_{\rm ph}=\hbar \omega_a \sum_k a^\dagger_k a_k$ with $\hbar\omega_a=67$\,meV being the energy of the out-of-plane phonons at K and K'. \cite{mohr_phonons}

In this model dealing with three electronic bands, the non-interacting electron Green function is a diagonal $3\times 3$ matrix and reads as
\begin{equation}
\label{eqn:G0}
G^0(q,i\omega_n)=\left(\begin{array}{ccc}\frac{1}{i\omega_n-E_\sigma}& 0 & 0\\
0 & \frac{1}{i\omega_n-\epsilon(q)} & 0 \\
0 & 0 & \frac{1}{i\omega_n+\epsilon(q)} \end{array}\right),\end{equation}
where $\omega_n$ are fermionic Matsubara frequencies.
With the non-interacting phonon Green function $D^0(k,i\Omega_m)=D^0(i\Omega_m)=-\frac{2\omega_a}{\Omega_m^2+\omega_a^2}$ and the electron-phonon interaction from Eqn. (\ref{eqn:V_e-ph}) transformed
to the matrix form of Eqn. (\ref{eqn:G0}), $\mathbf{M_-}=\left(\begin{array}{ccc}0&\lambda&0\\ \lambda&0&0 \\ 0 & 0&0 \end{array}\right)$ and  $\mathbf{M_+}=\left(\begin{array}{ccc}0&0&\lambda\\ 0&0&0 \\ \lambda & 0&0 \end{array}\right)$, the electronic self-energy is
%\begin{eqnarray}
%\Sigma(q,i \omega_n)&=&-\frac{1}{\beta}\sum_{k,\Omega_m,\nu}D(k,i\Omega_m)\mathbf{M}_\nu G^0(q-k,i\omega_n-i\Omega_m)\mathbf{M}_\nu\nonumber\\
%&=&-\frac{1}{\beta}\sum_{\Omega_m,\nu}D(i\Omega_m)\mathbf{M}_\nu G^0(r=0,i\omega_n-i\Omega_m)\mathbf{M}_\nu,
%\end{eqnarray}
\begin{equation}
\Sigma(i \omega_n)=-\frac{1}{\beta}\sum_{\Omega_m,\nu}D(i\Omega_m)\mathbf{M}_\nu G^0(r=0,i\omega_n-i\Omega_m)\mathbf{M}_\nu,
\end{equation}
where $G^0(r=0,i\omega_n)$ is the non-interacting
 local Green's function and the independence
  of $D^0$ and $\mathbf{M}_\nu$ of $k$ has been exploited.
   This self-energy is diagonal and, for energies $\omega\ll W$
   small as compared to the Dirac electron bandwidth $W\approx
   6$\,eV, its
   components read as $\Sigma_{1,1}(\omega+i\delta)=\Sigma'_{1,1}(\omega+i\delta)+i\Sigma''_{1,1}(\omega+i\delta)$,
where
\begin{widetext}
\begin{equation}
\Sigma'_{1,1}(\omega+i\delta)=\frac{2\lambda^2}{W^2}\left\{\begin{array}{ll}(\omega+\mu-\omega_a)\log\left|\frac{\omega-\omega_a}{W}\right|+(\omega+\mu+\omega_a)\log\left|\frac{(\omega+\omega_a+\mu)^2}{W(\omega+\omega_a)}\right| &\text{if}\;\mu\geq 0 \\
 (\omega+\mu-\omega_a)\log\left|\frac{(\omega+\mu-\omega_a)^2}{W(\omega-\omega_a)}\right|+(\omega+\mu+\omega_a)\log\left|\frac{\omega+\omega_a}{W}\right| & \text{if}\; \mu<0\end{array}\right. \nonumber\end{equation}
\end{widetext}
and
\begin{equation}
\Sigma''_{1,1}(\omega+i\delta)=-\frac{2\pi\lambda^2}{W^2}\Theta(|\omega|-\omega_a)\left|\omega+\mu-\sign(\omega)\omega_a\right|.
\end{equation}
The $\pi$ block is
\begin{equation}
\Sigma_{i,i}(\omega+i\delta)=\lambda^2\frac{1}{\omega+\mu-E_\sigma-\omega_a}\approx-\frac{\lambda^2}{E_\sigma},
\end{equation}
with $i=1,2$ and causes a rigid shift of the band energies in regime important for STS.

\begin{figure}
\includegraphics[width=.9\linewidth]{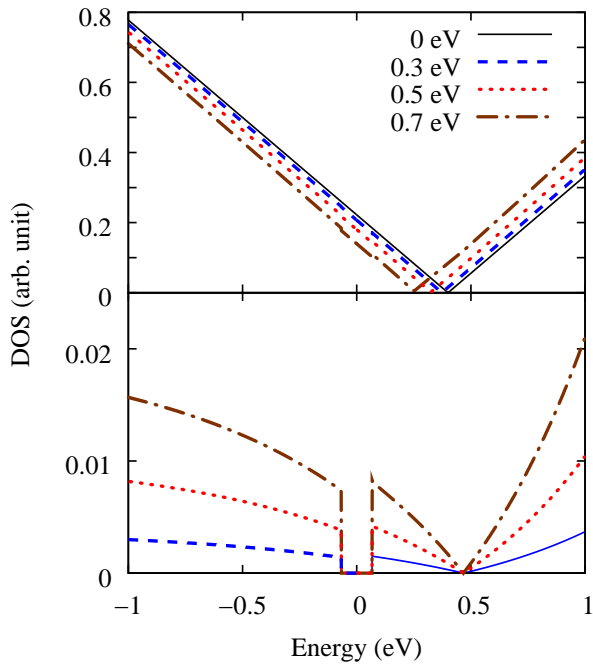}
\caption{\label{fig:DOS_all_sigma}(Color online) DOS for different coupling strengths $\lambda=0,\; 0.3,\; 0.5,\; 0.7$\, eV. Upper panel: Total DOS. Lower panel: DOS of nearly free electron band.}
\end{figure}

With the interacting Green function being $G^{-1}(p,\omega)={G^{(0)}}^{-1}(p,\omega)-\Sigma(\omega)$ momentum space integration yields the total density of states $N(\omega)=-\frac{1}{\pi}\Tr\;\Im G(r=0,\omega+i\delta)$.
For fixed chemical potential $\mu=-0.4eV$, the upper panel in Fig. \ref{fig:DOS_all_sigma} shows how the total DOS
is modified to the electron-phonon interaction as a function of the coupling strength $\lambda$. The total DOS and the DOS of the $\pi$-bands (not shown here) are virtually indistinguishable and exhibit the "V"-shape characteristic for Dirac fermions. The main modification of total DOS is a shift to slightly lower energies with increasing coupling strength due to the $\pi$-block of $\Sigma(\omega+i\delta)$ --- a consequence of level repulsion of the $\pi$ and the nearly free electron bands, well known to occur in second order perturbation theory.

However, the spectral properties of the nearly free electron channel are strongly altered at low energies $|\omega|\ll W$. The density of states in this channel reads as
\begin{eqnarray}
N_\Gamma(\omega)&=&-\frac{\Sigma''_{1,1}(\omega+i\delta)}{\pi|\omega-E_\sigma-\Sigma_{1,1}(\omega+i\delta)|^2}\nonumber\\
&\sim&\Theta(|\omega|-\omega_a)\left|\omega+\mu-\sign(\omega)\omega_a\right|.
\end{eqnarray}

Without electron-phonon interaction, the DOS of the nearly free electron states vanishes for $\omega< E_\sigma$. But, as soon as this interaction becomes effective and the electron's energy $\omega>\omega_a$ exceeds the phonon mode energy, the nearly free electron states start mixing with the $\pi$ bands. This leads to a gap of $\pm \omega_a$ around the Fermi level in the nearly free electron channel. Outside this gap, the nearly free electron DOS recovers the usual "V"-shape of graphene's DOS. (See Fig. \ref{fig:DOS_all_sigma}, lower panel.) Indeed, the shape of the DOS in this channel is very similar to the gapped spectra found in STS on graphene. \cite{Crommie_STS}

Comparing the scales in the upper and lower part of Fig. \ref{fig:DOS_all_sigma} one sees, that the DOS in the nearly free electron channel is for all coupling constants considered, here, smaller than the total DOS by a factor of less than 1/50. In STS, however, this factor of 1/50 is by far overcompensated by the factor of order $10^4$ coming from the different tunneling matrix elements in presence or absence of K/K' out of plane phonons.

\begin{figure}
\includegraphics[width=.9\linewidth]{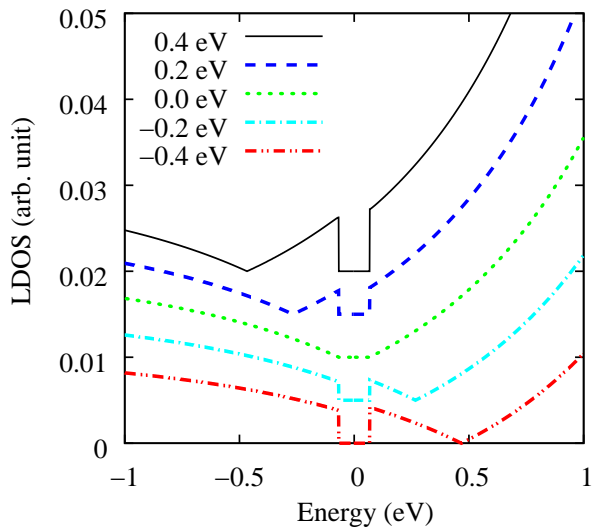}
\caption{\label{fig:DOS_sigma_mu}(Color online) DOS of nearly free electron states for different chemical potentials $\mu=-0.4,\; -0.2, \;0.0, \;0.2$ and $0.4$\,eV and the coupling strength $\lambda=0.5$\, eV fixed to the experimental value\cite{Crommie_STS}. The curves are vertically displaced for clarity..}
\end{figure}

As the STM spectra are dominated by the nearly free electron channel, it is illustrative to study $N_\Gamma$ as a function of the chemical potential, as shown in Fig. \ref{fig:DOS_sigma_mu}.
The gap of $\omega_a$ around the Fermi level is independent of doping, while the minimum caused by nearly free electron states coupled with $\pi$-states at the Dirac point shifts with the chemical potential. Experimentally, the chemical potential is shifted with the gate voltage \cite{Novoselov_science2004,Crommie_STS}. So, the spectra calculated, here, and the experimental STS from Ref. \cite{Crommie_STS} exhibit the same characteristic behavior.

The good agreement of the calculated tunneling DOS and the experimental STM spectra
firstly proves the correctness of the phonon mediated tunneling mechanism proposed
in Ref. \cite{Crommie_STS}. Moreover,
 it becomes clear that the phenomenon of strongly enhanced tunneling currents may
 occur in various contexts in graphene. No inelastic effects are required: The
  DFT calculations presented in the first part of this article have a static
  lattice and show the enhancement of the local DOS in vacuum in presence of
  frozen K/K' phonons. The only effect important for such an enhancement is
  the mixing of the Dirac electron states at the Brillouin zone corners K/K'
   with the nearly free electron states at $\Gamma$. Therefore, any out-of-plane
   corrugations on scale of the graphene lattice constant will increase the
   tunneling density of states significantly. This has important consequences
    for STM experiments: The "visibility" of impurity states at low
    energies\cite{wehling-2006-} will be strongly enhanced as soon as
    the impurity causes out of plane distortions of the graphene lattice
    STM images below the phonon threshold\cite{Rutter_Science} can be
     expected to look much more inhomogeneous than those taken at higher bias voltages.

The authors thank  E. Andrei, F. Binder, V. Brar,  M. F. Crommie, H.
Dahal, M. Galperin, J. Stroscio,  Y. Zhang, and J. X. Zhu for useful
discussions. This work was supported by US DOE at Los Alamos and SFB
668 (Germany). T.O.W. is grateful to LANL and the T11 group for
hospitality during the visit, when the ideas presented in this work
were conceived.
\bibliography{graphene}
\end{document}